\documentclass[prl, twocolumn, secnumarabic, amsmath, amssymb, natbib, superscriptaddress,nolongbibliography]{revtex4-2}
\usepackage{graphicx}
\usepackage{dcolumn}
\usepackage{bm}
\usepackage{float}
\usepackage{setspace}
\usepackage{multirow}   
\usepackage{amsmath}    
\usepackage{array}  
\usepackage{amssymb}             
\usepackage{amsfonts} 
\usepackage{microtype}
\usepackage{color}
\usepackage{xr-hyper}  
\newcommand{\Vratio}{V_{\mathrm{cell}}/V_{\mathrm{shell}}}
\newcommand{\Pe}{\mathrm{Pe}}

\begin{document}
\preprint{APS/123-QED}
\title{Confinement-Induced Symmetry Breaking of Active Surfaces}
\author{Da Gao}
\affiliation{Department of Physics, College of Physical Science and Technology, Xiamen University, Xiamen 361005, People’s Republic of China}
\author{Alexander Mietke}
\email{alexander.mietke@physics.ox.ac.uk}
\affiliation{Rudolf Peierls Centre for Theoretical Physics, Department of Physics, University of Oxford, Parks Road, Oxford OX1 3PU, United Kingdom}
\author{Rui Ma}
\email{ruima@xmu.edu.cn}
\affiliation{Department of Physics, College of Physical Science and Technology, Xiamen University, Xiamen 361005, People’s Republic of China}
\affiliation{Fujian Provincial Key Laboratory for Soft Functional Materials Research, Research Institute for Biomimetics and Soft Matter}

\begin{abstract}
The actomyosin cortex, a thin layer of a cross-linked polymer network near the cell surface, generates active forces that are responsible for cell shape changes. Many developmental processes that involve such cell shape changes, most prominently embryonic cell division, are spatially confined by eggshells. To investigate the potential role of confinement in redirecting active stresses and enabling symmetry breaking phenomena during cell shape transformations, we study a hydrodynamic minimal model in which the cell cortex is represented as an active fluid surface that undergoes symmetric division in the absence of confinement. When enclosed by an ellipsoidal shell, a spontaneous symmetry-breaking transition emerges at a critical degree of confinement, where symmetrically dividing surfaces become unstable and polarized geometries appear. We show that this transition is controlled by the tightness of the confinement and analyze the solution space of stationary surfaces to identify the mechanisms underlying confinement-induced symmetry breaking.
\end{abstract}
\maketitle
 Morphogenesis -- the emergence of shape in living systems -- relies not only on precisely orchestrated mechanical and biochemical processes, but is also often constrained by spatial confinement: Cells divide within crowded tissues, experience mechanical compression from neighbors, or are constrained by extracellular matrix~\cite{Evironment_cell}. Also the first cell divisions of most organisms that develop outside the parent's body are confined by some form of protective layer. In the canonical model organism \textit{C. elegans}, embryonic cell division occurs within a rigid, ellipsoidal eggshell~\hbox{\cite{Olson2012HierarchicalAO,PMID:32382445,Hsu2023ContractileRM,Johnston2006TheEI,Eggshell,MINC2011414}}. Yet the physical impact such confinement may have on  the cell shape control during division remains largely unexplored.

The cell cortex is a thin layer of cross-linked actin filaments and associated myosin motor proteins beneath the plasma membrane that surrounds cells. It plays a central role in cellular shape control~\cite{salbreux2012actin,Cotex_architecture,POLLARD2003453,Cytokinesis}. Myosin generates non-equilibrium stresses that drive cortical flows and give rise to contractile structures, such as the cytokinetic ring that pinches off a dividing cell~\hbox{\cite{2010Anisotropies,scienceFlow,Salbreux2009}}. During cell division, which occurs on minute time scales -- as compared to the actomyosin turnover that happens on second time scales~\cite{murth05,guha05} -- the cortex can be described as viscous material~\cite{ranft10} that harbors active stresses. Nonequilibrium hydrodynamic theories have been systematically constructed for such systems~\cite{marchetti2013hydrodynamics,Prost2015naturephysics,Julicher_2018}. Focusing on the role of surface-intrinsic active stresses, these theories successfully reproduce cortical flows~\cite{mayer2010anisotropies,naga14,pimp20} and cell shape transformations, including cell division and polarization~\cite{PhysRevLett.114.048102,PhysRevLett.128.068102,Turlier2014,PMID:22570593,PMID:21822289,DEKINKELDER2021110413,Mietke2019Self}. 

Here, we study how spatial confinement alters the dynamics of an active surface that undergoes cell-division-like shape transformations. To this end, we use a minimal active fluid surface model that exhibits robust symmetric divisions in the absence of confinement and consider an ellipsoidal enclosure that mimics an embryonic eggshell. Using a variational approach to determine the nonlinear solution space of deforming active surface models~\cite{gao2025}, we analyze the emerging shape dynamics as the dividing surface as confinement tightens. We show numerically and using analytic arguments that the degree of confinement serves as a control parameter to induce spontaneous symmetry breaking of a symmetrically ingressed active surface, leading to geometrically polarized stationary and oscillatory states. Our analysis reveals that confinement stabilizes solution branches with polar surface geometries, which have a shorter pole-to-pole distance than symmetrically constricted surfaces and therefore can more easily accommodate tight confinement.

We model the cell cortex as an axisymmetric surface parameterized by 
\begin{equation}
    \mathbf{X}(u,\phi,t)=[r(u,t)\cos\phi,\; r(u,t)\sin\phi,\; z(u,t)],
\end{equation}
where $u \in [0,1]$ and $\phi \in [0,2\pi]$ denote surface coordinates. The tangent basis vectors are $\mathbf{e}_i=\partial_i \mathbf{X}$ with $i\in\{u,\phi\}$, and the unit normal is $\mathbf{n}=(\mathbf{e}_u \times \mathbf{e}_\phi)/|\mathbf{e}_u \times \mathbf{e}_\phi|$. A general velocity field decomposes into tangential and normal components, $\mathbf{v}=\mathbf{v}_{\parallel}+\mathbf{v}_{\perp}=v_i\mathbf{e}^i+v_n\mathbf{n}$, where $\mathbf{v}_{\parallel}$ describes in-plane flows and $\mathbf{v}_{\perp}$ describes deformations. The surface dynamics and local center of mass velocity field $\mathbf{v}$ are related by 
\begin{equation}
    \frac{\mathrm{d}\mathbf{X}}{\mathrm{d}t}=\mathbf{v},
\end{equation}
where $\mathrm{d}/\mathrm{d}t(\cdot)$ denotes the total time derivative. The geometry of the surface is characterized by the metric tensor $g_{ij}=\mathbf{e}_i\cdot\mathbf{e}_j$ and the curvature tensor \hbox{$C_{ij}=-\mathbf{n}\cdot\partial_i\partial_j\mathbf{X}$}. The symmetric part of the velocity gradient tensor is $S_{ij}=({\nabla_i}v_j+{\nabla_j}v_i)/2+C_{ij}v_n$, and we denote its traceless part by $\tilde{S}_{ij}=S_{ij}-\tfrac{1}{2}S_k^kg_{ij}$~\cite{Guillaume2017}.

\begin{figure}
    \includegraphics[width=1\columnwidth]{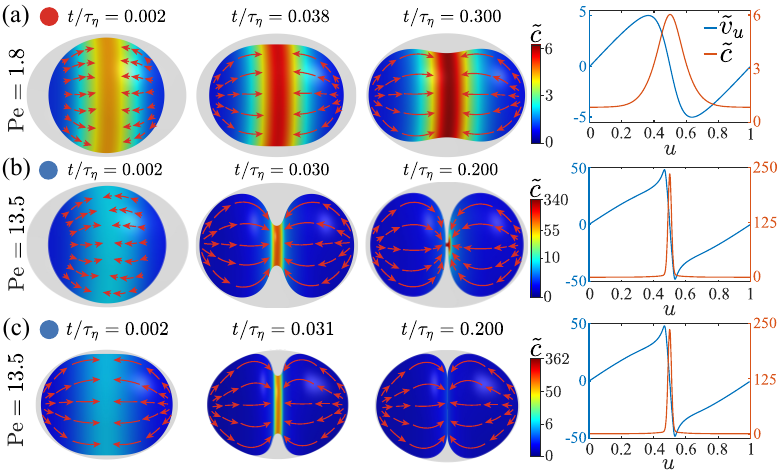}
  \caption{
\textbf{Confined symmetric divisions.}
(a)~Surface dynamics (Movie~1) illustrating tangential surface flows $\tilde{v}_u=\bar{v}_u\tau_{\eta}/R_0$ (red arrows) and the normalized stress-regulator concentration $\tilde{c}=c/c_0$ under weak confinement ($\epsilon=0.64$, where $\epsilon=\Vratio$) and weak contractility ($\mathrm{Pe}=1.8$). (b)~Same confinement as in (a), but with increased activity ($\mathrm{Pe}=13.5$) (Movie~2). (c)~Dynamics with strong confinement ($\epsilon=0.87$) at the same contractility as in (b) ($\mathrm{Pe}=13.5$) (Movie~3). The transparent gray surface illustrates the confining shell. Right panels show profiles of $\tilde{v}_u$ and $\tilde{c}$ of the final steady states. All other parameters are listed in Table~\ref{tab:para_surface}.}
    \label{fig:Shell_effect}
\end{figure}

The surface dynamics are governed by the force balance~equation
\begin{equation}
\label{eq:forcebalance}
    \text{div}(\mathbf{T}) = -\mathbf{f}^{\mathrm{ext}},
\end{equation}
where $\mathbf{T}=t_{ij}\mathbf{e}^{i}\!\otimes\!\mathbf{e}^j+t_{i,n}\mathbf{e}^i\!\otimes\!\mathbf{n}$ denotes the total surface stress tensor and $\mathbf{f}^{\mathrm{ext}}=f^{\text{ext}}_i\mathbf{e}^i+f^{\text{ext}}_n\mathbf{n}$ the external force density. The surface divergence operator is defined as $\text{div}(\cdot)=\mathbf{e}^i\cdot\partial_i(\cdot)$. Following previous work on minimal models of active self-organized surfaces~\cite{Mietke2019Minimal,gao2025}, we consider a surface stress tensor that comprises of both equilibrium and dissipative contributions. The former~read
\begin{align}
\label{eq:elastic tension tensor}
    t_{ij}^e&=\frac{\kappa}{2} C^k_{\;k}\!\left(C^{k}_{\;k}g_{ij}-2C_{ij}\right)\\
\label{eq:tne}
    t_{i,n}^{e}&=\kappa\nabla_i C^k_{\,k}
\end{align}
and follow from a Helfrich bending energy~\cite{Helfrich1973,RCapovilla_2002,Guillaume2017}. The dissipative tension tensor is chosen as~\cite{Mietke2019Minimal,Mietke2019Self,gao2025}
\begin{equation}
\label{eq:dis}
    t^d_{ij} = 2\eta_s \tilde{S}_{ij} + \eta_b \,\mathrm{div}(\mathbf{v})\, g_{ij} + \xi \Delta \mu\, g_{ij},
\end{equation}
where $\eta_s$ and $\eta_b$ denote the surface shear and bulk viscosities, respectively. The last term describes an isotropic active tension with contractility $\xi$ and chemical potential difference~$\Delta\mu$.

The in-plane components of the external force density~$\mathbf{f}^{\text{ext}}$ are given by
\begin{equation}\label{eq:fext}
f^{\text{ext}}_j=-\Gamma v_j,
\end{equation}
where $\Gamma$ denotes an effective friction coefficient. In this work, we consider an external normal forces given by
\begin{equation}\label{eq:fn}
f^{\text{ext}}_n=p+f^{c}_n.
\end{equation}
In addition to a pressure $p$ difference across the surface that enforces conservation of the enclosed volume, we include in this work a force density $f_n^c$ in Eq.~(\ref{eq:fn}) that imposes a spatial confinement of the surface into a prolate ellipsoidal volume $V_{\text{shell}}$. We implement this force using a potential
\begin{equation}
\label{eq:confine_potential}
P^{\text{ext}}=P_0\exp\left[\frac{1}{{w_p}^2}\left(\frac{z^2}{a^2}+\frac{r^2}{b^2}-1\right)\right],
\end{equation}
which is constant on ellipsoidal surfaces with $a$ and $b$ being the semi-major and minor axes ($a\ge b$), and $V_{\text{shell}}=(4\pi/3)ab^2$. The confining force is then given by
\begin{equation}
f_n^c =- (\nabla P^{\text{ext}})\cdot \mathbf{n}.
\end{equation}
We choose $w_p$ small enough that the confining potential exerts a significant force only near the boundary, and $P_0$ large enough to ensure that the surface shape remains within the confinement. Balancing numerical stability and physical fidelity, we use $w_p = 0.14$ and $P_0 = 0.9 \kappa/R_0^2$.


\begin{figure*}
    \includegraphics[width=2.05\columnwidth]{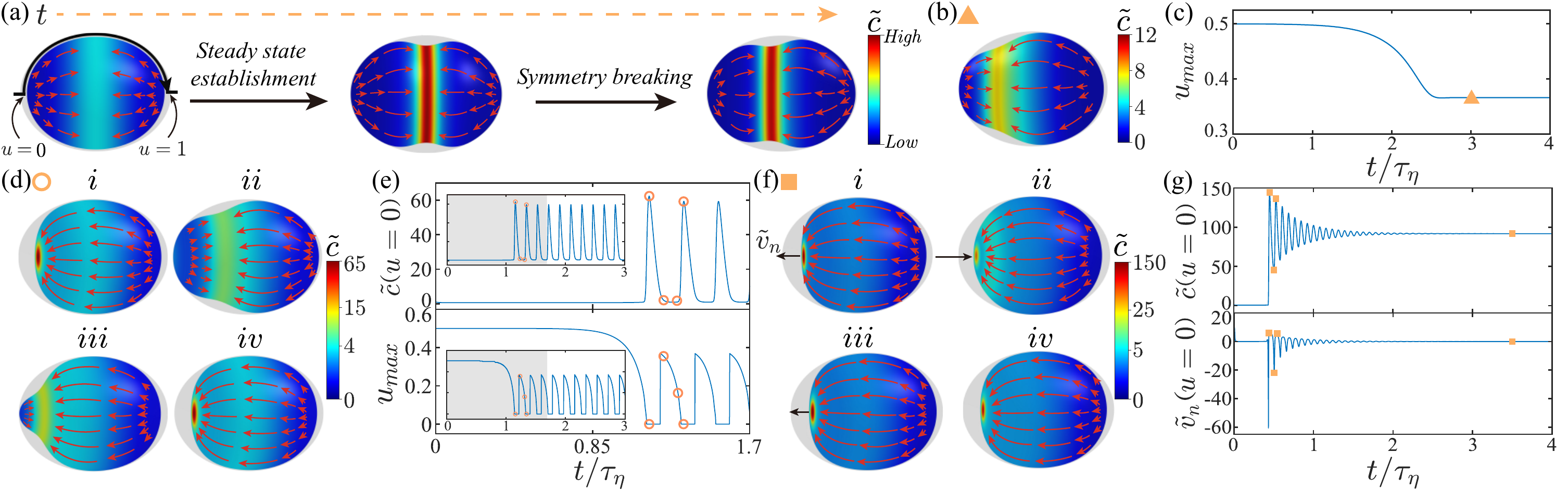}
    \caption{
\textbf{Dynamics of confinement-induced symmetry breaking.}
(a)~Representative evolution of a symmetry-breaking event showing stress regulator concentration $\tilde{c}=c/c_0$ (color scale) and tangential surface flows $\tilde{v}_u$ (red arrows).
(b)~Weakly polarized steady state in which the contractile ring stabilizes at an intermediate position between the pole and the equatorial plane~($\mathrm{Pe}=3.83$, Movie~4).
(c)~Time evolution of the location of maximal stress regulator concentration~$u_{\mathrm{max}}(t)$. Orange triangle depicts the time corresponding to snapshot in (b).
(d)~One cycle of a persistent oscillatory state of asymmetric contractile ring formation and slipping~($\mathrm{Pe}=4.5$, Movie~5).
(e)~Temporal evolution of stress regulator concentration $\tilde{c}$ at pole $u=0$ (top) and $u_{\mathrm{max}}$ (bottom) for dynamics shown in (d). (f) One damped oscillation cycle ($i-iii$) followed by the final steady state ($iv$). ($\mathrm{Pe}=6.3$, Movie~6).
Black arrows indicate magnitude of the normal velocity $\tilde{v}_n$ at the pole ($u=0$).
(g)~Temporal evolution of $\tilde{c}$ (top) and the normal velocity $\tilde{v}_n$ (bottom) at $u=0$ associated (f). Data shown uses confinement magnitude $\epsilon=0.87$, all other parameters are listed in Table~\ref{tab:para_surface}~\cite{SItext}. 
}\label{fig:Oscillation}
\end{figure*}

Active tension in Eq.~(\ref{eq:dis}) is regulated by the concentration field $c$ of a stress regulator as~\cite{Bois2011PRL,Kumar2014PRL} 
\begin{equation}
   \xi \Delta\mu = (\xi \Delta\mu)_0\, H_+(c)
   = (\xi \Delta\mu)_0 \frac{c^2}{c^2 + c_s^2},
\end{equation}
where $c_s$ denotes the saturation concentration. The dynamics of the regulator concentration are governed by a continuity equation
\begin{equation}
\label{eq:dcdt}
        \frac{\mathrm{d}c}{\mathrm{d}t}
        = -\,c\,\mathrm{div}(\mathbf{v})
          + D\,\Delta_{\Gamma} c
          - k\,(c - \hat{c}_{0}),
\end{equation}
where terms on the right capture local dilation and compression effects, in-plane diffusion characterized by a diffusion constant $D$ with Laplace--Beltrami operator $\Delta_{\Gamma} c = \nabla^{i}\nabla_{i} c$, and the last term describes exchange with the surrounding in terms of an off-rate $k$ and a steady-state concentration $\hat{c}_0$. For uniform $\hat{c}_0$ this model predicts -- beyond a critical contractility $\xi$ -- a spontaneous surface polarization in which stress regulator accumulates at one pole~\cite{Mietke2019Self,Mietke2019Minimal,gao2025}. To capture the mechano-chemical features of a division-like dynamics instead, we therefore prescribe $\hat{c}_0(u)$ as a Gaussian function peaked at the surface midplane:
\begin{equation}\label{eq:c0prof}
    \hat{c}_{0}(u)
    = c_0 \!\left[
        1 + \Delta_c
        \exp\!\left(
            -\frac{\left(u-\tfrac{1}{2}\right)^2}{w^2}
        \right)
    \right],
    \qquad (\Delta_c > 0).
\end{equation}
This functional form parametrizes upstream regulatory mechanisms that preferentially recruits stress regulator to the equatorial region, consistent with experimental observations of spindle--cortex interactions~\cite{ther07,neli15,dead23}. Parameters~$\Delta_c$ and~$w$ in Eq.~(\ref{eq:c0prof}) set the regulation strength and width of the regulated region, respectively.


Following~\cite{gao2025}, we convert force balance and continuity Eqs.~(\ref{eq:forcebalance}) and~(\ref{eq:dcdt}) into a boundary value problem (BVP) that can be solved iteratively to determine the time evolution of the surface shape $\mathbf{X}$, the velocity field~$\mathbf{v}$, and the concentration field $c$~\cite{SItext}. The dynamics are initialized with a homogeneous concentration $c=c_0$, a vanishing flow field $\mathbf{v}=0$, and a surface shape~$\mathbf{X}$ that is stationary in the absence of contractility ($\xi\Delta\mu=0$) for the given confinement. Characteristic time and length scales are defined as $\tau_{\eta} = \eta_{\mathrm{b}} R_0^2 / \kappa$ and $R_0$, respectively, where the cell volume is $\mathrm{V_{cell}}=4\pi R_0^3/3$. The magnitude of active contractility is characterized by the Péclet number $\mathrm{Pe}=(\xi\Delta\mu)_0 H_+(c_0)R_0^2/(D\eta_{\mathrm{b}})$. A key parameter in our analysis is the degree of confinement, $\epsilon=\Vratio$, which ranges from 0 (no confinement, $\mathrm{V_{shell}}\rightarrow\infty$) to 1 for maximally tight confinement.

We first demonstrate several representative dynamical behaviors associated with symmetric division. Under relatively weak confinement ($\epsilon = 0.64$) and weak contractility ($\mathrm{Pe} = 1.8$), the stress regulator accumulates at the equator and forms a contractile ring that drives surface flows from the poles toward the equator. As a result, the surface eventually adopts a partially ingressed steady state shape~(Fig.~\ref{fig:Shell_effect}a). As contractility is increased, surface flows gain magnitude, and the stress regulator becomes even more strongly enriched at the equator. Consequently, final steady states exhibit more and more pronounced constrictions. At strong contractility ($\mathrm{Pe} = 13.5$), the final steady state surface is almost completely ingressed, resulting in a $\mathrm{V}$-shaped geometry near the equator~(Fig.~\ref{fig:Shell_effect}b). Tightening the confinement ($\epsilon = 0.87$) still allows for ingression, but daughter cells remain very close throughout the division process, and hardly any space is left between them at the final steady state~(Fig.~\ref{fig:Shell_effect}c). Such dynamics and cell morphology closely resemble that of the \textit{C.~elegans} zygote during its first embryonic division~\cite{eLife}.

\begin{figure*}
    \includegraphics[width=2\columnwidth]{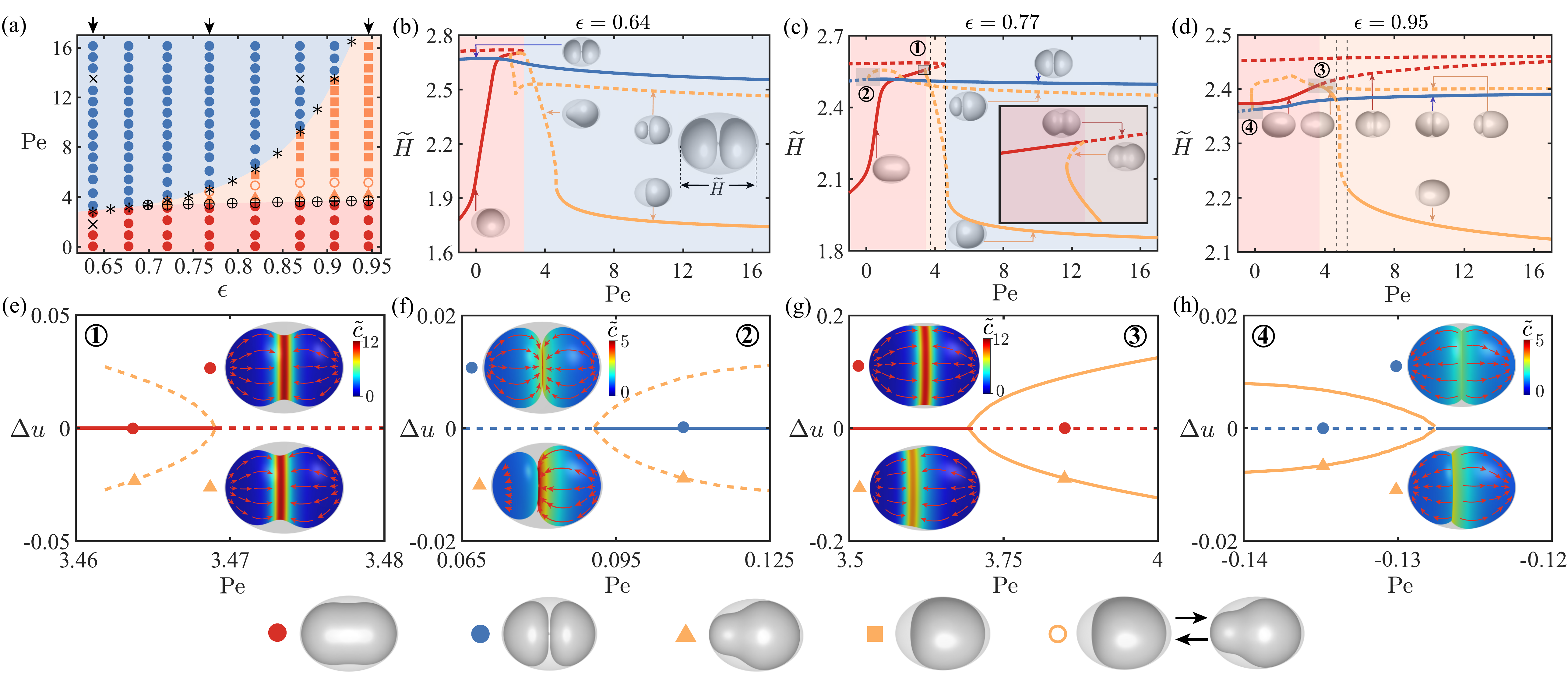}
    \caption{
\textbf{Phase diagram and stationary shape spaces of confined active surfaces.}(a)~Phase diagram for the degree of confinement $\epsilon=\Vratio$ and the P\'eclet number $\mathrm{Pe}=(\xi\Delta\mu)_0 H_+(c_0)R_0^2/(D\eta_{\mathrm{b}})$ quantifying the strength of activity. Symbols depict numerical simulations (representative shapes at the bottom). Background shades indicate surface geometry: symmetric weakly (red) and strongly (blue) ingressed, and polarized (orange). Parameters used in Fig.~\ref{fig:Shell_effect} are indicated by $\times$. Turning points of solution branches formed by symmetric, weakly ingressed surfaces (red lines in panels b--d) are denoted by $*$. Numerically determined pitchfork bifurcation points ($+$) agree well with critical points predicted from semi-analytical arguments ($\circ$, see Eq.~\eqref{eq:growthrate} and Fig.~\ref{fig:Pe*_Tran} in \cite{SItext}). (b--d)~Pole-to-pole distance $\tilde{H}=H/R_0$ along stationary solution branches for different confinement degrees (black arrows in a). Solid (dashed) lines indicate stable (unstable) solutions and color shading as in panel a. Transitions to polarized surfaces (red to orange) and to strongly ingressed symmetric surfaces (red/orange to blue) in the phase diagram are determined by bifurcations and turning points, respectively, along symmetric, weakly ingressed surface branches (red lines). Oscillations occur for Pe bound by dashed lines in (c) and (d). Polar surfaces generally exhibit shorter pole-to-pole distances and can therefore more easily accommodate tight confinement. (e--h)~Solution branches of weakly (e) and strongly (f) ingressed symmetric surfaces connect to corresponding polar surfaces via subcritical bifurcations (e,f) that become supercritical (g,h) when tightening the confinement. $\Delta u = u_{\mathrm{max}} - \tfrac{1}{2}$ quantifies the asymmetry of the contractile ring position and scales as $|\mathrm{Pe} - \mathrm{Pe}^*|^{1/2}$ in the vicinity of the bifurcation point (see Fig.~\ref{fig:Bifur_varAc_Root} in \cite{SItext}).}\label{fig:PhaseCurve}
\end{figure*}

At intermediate P\'{e}clet numbers, confinement induces a spontaneous symmetry breaking: the contractile ring ceases to constrict symmetrically and instead slips toward one of the poles (Fig.~\ref{fig:Oscillation}a). Depending on the value of~$\mathrm{Pe}$, three distinct regimes emerge from this ring slip. We characterize these regimes by tracking the position $u\in[0,1]$ of the concentration maximum over time and denote the resulting function by $u_{\mathrm{max}}(t)$. For increased~$\mathrm{Pe}$, the slipped ring remains at a position between a pole and the equator, leading to a stationary, asymmetrically ingressed pear-shaped geometry~(Fig.~\ref{fig:Oscillation}b,c, $\mathrm{Pe} = 3.83$).  Increasing $\mathrm{Pe}$ further, the ring fully slips and a cluster of high stress regulator concentration remains at the pole. Two distinct behaviors were then observed: (1)~\textit{Persistent oscillations}, where the polar cluster dissolves and a new ring reassembles at an intermediate position before migrating poleward again to repeat the cycle (Fig.~\ref{fig:Oscillation}d,e, $\mathrm{Pe} = 4.5$); and (2)~\textit{Damped oscillations}, where the pole towards which the ring slipped exhibits damped shape oscillations and eventually remains flattened due to locally increased stress regulator concentration~(Fig.~\ref{fig:Oscillation}f,g,~$\mathrm{Pe} = 6.3$).
 
A general overview of the parameter regions that give rise to these dynamical regimes is provided in the phase diagram~Fig.~\ref{fig:PhaseCurve}(a), which is spanned by the confinement degree $\epsilon$ and the P\'{e}clet number $\mathrm{Pe}$.
At weak confinement ($\epsilon \lesssim 0.7$), surfaces relax to symmetrically ingressed stationary geometries: a partially ingressed shape at low $\mathrm{Pe}$ (Fig.~\ref{fig:PhaseCurve}a, red shade) and a completely ingressed morphology at high $\mathrm{Pe}$ (Fig.~\ref{fig:PhaseCurve}a, blue shade).
As confinement  becomes tighter ($\epsilon \gtrsim 0.7$), stationary surfaces with polar symmetry emerge (orange shade in Fig.~\ref{fig:PhaseCurve}a; see Fig.~\ref{fig:Oscillation}b, d, f for representative examples), amounting to a confinement-induced symmetry breaking and polarization. This region spans an increasingly large range of P\'{e}clet numbers as the confinement $\epsilon$ is tightened further.
 

To rationalize the mechanism of confinement-induced symmetry breaking, we directly compute stationary geometries as a function of $\mathrm{Pe}$ and systematically evaluate their mechanical stability for various degrees of confinement. The results are depicted in Fig.~\ref{fig:PhaseCurve}b-d, where solid (dashed) lines indicate stable (unstable) solutions. At weak confinement (Fig.~\ref{fig:PhaseCurve}b, $\epsilon = 0.64$), we find for a passive fluid surface ($\mathrm{Pe} = 0$) two stable stationary surface geometries: An expected prolate ellipsoidal surface that reflects the confinement geometry (red solid curve at $\mathrm{Pe} = 0$), and surprisingly, also an almost completely ingressed surface geometry (blue solid curve at $\mathrm{Pe} = 0$) -- a solution that does not exist in the absence of confinement~\cite{gao2025}. Therefore, confinement can stabilize a division-like surface geometry even for passive surfaces. Increasing the Péclet number $\Pe$, prolate surfaces ingress further and the pole-to-pole distance $\mathrm{H}$ increases due to volume conservation~(Fig.~\ref{fig:PhaseCurve}b-d, solid red lines). For weak confinement (Fig.~\ref{fig:PhaseCurve}b), this branch is stable up to a turning point, where the surface reaches a maximum pole-to-pole extension $\mathrm{H_{max}}\approx2a$, as dictated by the confining ellipsoid [see Eq.~(\ref{eq:confine_potential})]. In contrast,~$\mathrm{H}$ decreases monotonically on the branch of symmetric completely ingressed surfaces when increasing $\Pe$~(Fig.~\ref{fig:PhaseCurve}b-d, solid blue lines). Not only in the absence of confinement~\cite{gao2025}, but also for finite very weak confinement ($\epsilon\lesssim0.4$), weakly and strongly ingressed branches are connected by a Gibbs loop and fully ingressed geometries are only stable for $\text{Pe}>0$~(see Fig.~\ref{fig:top_Tran} in~\cite{SItext}). As $\epsilon$ increases and confinement tightens, this Gibbs loop deforms, and fully ingressed, stable geometries can also exist for $\text{Pe}=0$.

Beyond these symmetric regimes, we identify a third set of solution branches along which surface geometries obtain a polar asymmetry (Fig.~\ref{fig:PhaseCurve}b-d, orange lines). These branches encompass two kinds of scenarios. One is characterized by the displacement of the contractile ring to an intermediate position between the pole and the equatorial plane, which we refer to as \textit{weakly polarized surfaces}. Depending on the magnitude of contractile tension, the resulting geometry is either pear-shaped or is characterized by an asymmetrically positioned, strongly ingressed furrow (inset shapes in Fig.~\ref{fig:PhaseCurve}b show representative examples). The second type of polar surface geometry -- \textit{strongly polarized surfaces} —appears when the contractile ring has fully slipped, while a cluster of high stress regulators remains at a significantly flattened pole. Overall, we find that weakly polarized surfaces are only stable in very narrow P\'eclet number intervals, while strongly polarized surfaces are stable for a broader range of Pe (Fig.~\ref{fig:PhaseCurve}a, see orange triangles and squares).

Tightening the confinement leads to significant changes in the stationary shape space~(Fig.~\ref{fig:PhaseCurve}c, $\epsilon=0.77$). In contrast to the weaker confinement (Fig.~\ref{fig:PhaseCurve}b), the stable part of the branch of symmetric weakly constricted surfaces (red solid line) now becomes unstable before it reaches the turning point imposed by the confinement (Fig.~\ref{fig:PhaseCurve}c, inset). At the bifurcation, a weakly polarized branch emerges via a subcritical pitchfork bifurcation (Fig.~\ref{fig:PhaseCurve}e). The pole-to-pole distance $\mathrm{H}$ \textit{decreases} along this new branch, and surfaces become stable within a narrow range of P\'{e}clet numbers (Fig.~\ref{fig:PhaseCurve}c, inset). This marks the initial appearance of stable surface geometries with polar asymmetry in the phase diagram (Fig.~\ref{fig:PhaseCurve}a) and suggests a mechanism for confinement-induced symmetry breaking: Geometric polarization of weakly constricted surfaces reduces their pole-to-pole distance, making it easier to adapt to tight confinement and leading to the destabilization of symmetric surfaces with larger pole-to-pole distances. At high~$\Pe$, stable strongly polarized surface geometries generally appear on this branch for all magnitudes of confinement (Fig.~\ref{fig:PhaseCurve}b-d, solid orange lines), corresponding to final states of the damped oscillations shown in Fig.~\ref{fig:Oscillation}f. For $\Pe$ beyond, but close to, the symmetry breaking transition (P\'eclet number intervals bounded by black dashed lines in Fig.~\ref{fig:PhaseCurve}c,d), neither weakly nor strongly polarized geometries are stable. As a result, persistent oscillations between these two states (Fig.~\ref{fig:Oscillation}d) emerge from a transition that exhibits signatures of a Hopf bifurcation~\cite{SItext}, i.e. an oscillation frequency that scales as $\omega \propto |\mathrm{Pe}-\mathrm{Pe}^*|$ and an oscillation amplitude that scales as $A \propto |\mathrm{Pe}-\mathrm{Pe}^*|^{1/2}$~(see Fig.~\ref{fig:Oscillation_fit} in~\cite{SItext}).
 
For sufficiently tight confinement, a new bifurcation also appears on the branch of fully ingressed surfaces~(Fig.~\ref{fig:PhaseCurve}\hbox{c-d}, blue lines). Analyzing first the case $\epsilon=0.77$ (Fig.~\ref{fig:PhaseCurve}c), a subcritical pitchfork bifurcation appears at small but finite Pe~(Fig.~\ref{fig:PhaseCurve}f) and gives rise to an unstable branch of weakly polarized, strongly ingressed surface geometries (dashed orange lines). For very tight confinement (Fig.~\ref{fig:PhaseCurve}d, $\epsilon=0.95$), the two bifurcations illustrated in Fig.~\ref{fig:PhaseCurve}e (weak ingression) and Fig.~\ref{fig:PhaseCurve}f (strong ingression) undergo a criticality transition from subcritical to supercritical (Fig.~\ref{fig:PhaseCurve}g,h). The bifurcation shown in Fig.~\ref{fig:PhaseCurve}h implies for fully ingressed surface geometries that, under extensile active tension $\xi<0\Rightarrow\Pe<0$, symmetric surfaces can become unstable and give way to stable surfaces with polar asymmetry~(Fig.~\ref{fig:PhaseCurve}h).

The loss of stability of symmetric surfaces at bifurcations (Fig.~\ref{fig:PhaseCurve}e and g), as well as the presence of stable weakly polarized surface geometries at or nearby these bifurcations, delineates the boundary between symmetric (blue) and polarized (orange) surface geometries in Fig.~\ref{fig:PhaseCurve}a. Notably, the critical Péclet number~$\mathrm{Pe}^*$ associated with this symmetry-breaking transition seems insensitive to the degree of confinement $\epsilon$. To rationalize this, we analyze steady-state concentration and flow profiles along the branch of weakly ingressed symmetric surfaces (red lines). Stress-regulator concentration is strongly localized at the equatorial plane ($u=1/2$), where it reaches its maximum value $c_{\mathrm{max}}$, while the meridional flow $v^u$ exhibits a significant gradient $\partial_u v^u$. A linear stability analysis yields a semi-analytical expression for the effective growth rate $\lambda$ of surface perturbations~\cite{SItext},
\begin{equation}
\label{eq:growthrate}
    \lambda \approx -D q_{\mathrm{min}}^2 - k - \partial_u v^u + \frac{\xi\Delta\mu}{\eta_b} H_{+}'(c_{\mathrm{max}}) c_{\mathrm{max}},
\end{equation}
where $q_{\mathrm{min}}$ is a fitting parameter that denotes the minimum wavenumber to be considered for linear stability analysis. While diffusive damping ($-D q_{\mathrm{min}}^2$) and degradation ($-k$) stabilize a symmetric surface, the active stress and flows promote shape polarization. Values $\Pe^*$ predicted with Eq.~(\ref{eq:growthrate}) agree well with numerically determined bifurcation points (compare black $+$ and $\circ$ in Fig.~\ref{fig:PhaseCurve}a). This analysis also reveals that the instability is primarily driven by compressive flow gradients ($-\partial_u v^u$) at the equator as contractility $\Pe$ increases~\cite{SItext}. The turning points of the symmetric, weakly ingressed branch delineate the phase boundary between the polarized (orange) and symmetric division (blue) regimes in the phase space (Fig.~\ref{fig:PhaseCurve}a). For values of Pe beyond these turning points, the stationary shape space no longer holds stationary solutions of symmetric, weakly ingressed surfaces. Instead, a sufficiently large active contractility (i.e. large $\Pe$) suppresses the symmetry-breaking instability and drives surfaces again toward a symmetric, completely ingressed geometry.

In this Letter, we have demonstrated that spatial confinement can destabilize symmetrically constricting active fluid surfaces and give rise to a spontaneously emerging polar asymmetry. Mathematically, this symmetry-breaking arises from confinement-enhanced pitchfork bifurcations in the stationary shape space. Physically, it arises because constricted surfaces with polar asymmetry fit more easily into tight confinement than symmetric~ones. 

Asymmetric division is a hallmark of early \textit{C.~elegans} development and is linked to a pre-established biochemical polarity ~\cite{spindle1,spindle2}. The degree of confinement during this division is approximately $\epsilon=0.85$~\cite{PhysRevLett.114.048102,TangZhang,gessele2020}, a value that resides within the polarized regime of our stationary shape space for a broad range of $\Pe$ (Fig.~\ref{fig:PhaseCurve}a, orange shaded region). Our results suggest that spatial confinement provided by the eggshell could play a critical role in reinforcing this asymmetry. This is also consistent with experimental observations of more symmetric divisions when the eggshell surrounding \textit{C. elegans} embryos is removed~\cite{ReShell}.  Our analysis also indicates that asymmetric, strongly ingressed surfaces are mechanically unstable under contractile active tension, which suggests that additional mechanical factors are required to sustain complete asymmetric divisions. Candidates for this missing ingredient are anisotropic active tension~\cite{Salbreux2009,reym16} and spontaneous curvature~\cite{dead23}. Importantly, cells that divide within biological tissues also experience effective confinement due to surrounding cells and the extracellular matrix. Motivated by these observations, it will be interesting to investigate how biochemical polarity cues and confinement with alternative geometric and mechanical properties influence the shape dynamics and robustness of cell division.



We acknowledge financial support from the National Natural Science Foundation of China under Grant No. 12474199 and the Fundamental Research Funds for Central Universities of China under Grant No. 20720240144, as well as from the 111 project B16029.

\appendix
\renewcommand{\theequation}{S\arabic{equation}}
\setcounter{equation}{0}
\renewcommand{\thefigure}{S\arabic{figure}} 
\renewcommand{\thetable}{S\arabic{table}} 
\setcounter{figure}{0} 
\setcounter{table}{0} 
\onecolumngrid
\section{Numerical approach}\label{sec:NumApr}
The numerical approaches used in this work are described in detail in ref.~\cite{gao2025} and the corresponding code is available on GitHub as referenced therein. To be self-contained, we provide in the following a brief summary.

\subsection{Parameterized axisymmetric surfaces}

\label{AP:Geo}
We consider deforming axisymmetric surfaces parameterized as
\begin{equation}\label{eq:para}
\mathbf{X}(u,\phi,t)=\big[r(u,t)\cos\phi,\; r(u,t)\sin\phi,\; z(u,t)\big],
\end{equation}
where $r(u,t)$ and $z(u,t)$ denote the time-dependent meridional profiles of the surface coordinates, and we refer to the time-independent parameterization $u\in[0,1]$ and $\phi\in[0,2\pi]$ as mesh coordinates (Fig.~\ref{fig:scheme}). Tangent vectors $\mathbf{e}_i=\partial_i\mathbf{X}$ and surface normal $\mathbf{n}=\mathbf{e}_u\times\mathbf{e}_{\phi}/|\mathbf{e}_u\times\mathbf{e}_{\phi}|$ are given by
\begin{equation}\label{eq:bvs}
\mathbf{e}_u=h\bar{\mathbf{e}}_u=h\begin{pmatrix}
\cos\psi\cos\phi \\
\cos\psi\sin\phi \\
-\sin\psi
\end{pmatrix}
\quad
\mathbf{e}_{\phi}=r\bar{\mathbf{e}}_{\phi}=r\begin{pmatrix}
-\sin\phi\\
\cos\phi \\
0
\end{pmatrix}\\
\quad
\mathbf{n}=\begin{pmatrix}
\sin\psi\cos\phi \\
\sin\psi\sin\phi \\
\cos\psi
\end{pmatrix},
\end{equation}
where $h=\sqrt{(r')^2+(z')^2}$ is a scaling factor ($'$ denotes derivative w.r.t. $u$), $\psi$ is the tangent angle, and $\bar{\mathbf{e}}_u$ and $\bar{\mathbf{e}}_{\phi}$ are normalized tangent vectors. Scaling factor $h$ and tangent angle $\psi$ and scaling factor are related to surface coordinates~by
\begin{align}
r' &= h\cos\psi, \label{eq:dur}\\
z' &= -h\sin\psi,
\label{eq:duz}
\end{align}
where primes denote derivatives with respect to $u$. Metric tensor $g_{ij}=(\partial_i\mathbf{X})\cdot(\partial_j\mathbf{X})$ and area element $dA=\sqrt{\mathrm{det}(g_{ij})}ds^1ds^2$ read
\begin{align}
g_{ij}&=\begin{pmatrix}
h^2 & 0\\
0 & r^2
\end{pmatrix}\label{eq:gijaxi}\\
dA&=hrdud\phi.
\end{align}

Curvature tensor components $C_i^{\,j}=-g^{jk}\mathbf{n}\cdot\nabla_i\nabla_k\mathbf{X}$ are given by
\begin{align}
C_i^{\,j}
&
=
\begin{pmatrix}
\psi'/h & 0 \\
0 & \sin\psi/r
\end{pmatrix},
\label{eq:Cijaxi}
\end{align}
where the diagonal elements correspond to the two principal curvatures of the surface.

\subsection{Constrained dissipation functional}
We solve the force balance Eq.~(3) (main text) and evolve the surface according to Eq.~(2) using a dissipation functional formulation (see \cite{gao2025} for details). This functional assumes  the form
\begin{equation}
\label{eq:FinalRay}
\begin{split}
\mathcal{R}
&=
2\pi\int_0^1
\Bigg[
\partial_t f_\kappa
+ f_\kappa\,\frac{\partial_t(rh)}{rh}
- r_p\,\Delta\mu
- \mathbf{f}_c^{\mathrm{ext}}\!\cdot\mathbf{v}
+ \frac{1}{2}\left(S_1+S_2+S_3\right)
+ \frac{1}{2}\Gamma\left(\bar{v}_u^2+\bar{v}_\phi^2\right)
\Bigg]
r h\, du \\
&\quad
+ 2\pi\int_0^1 du\,
\partial_t\!\left[
\alpha\left(r'-h\cos\psi\right)
+\beta\left(z'+h\sin\psi\right)
+\zeta\, h'
\right],
\end{split}
\end{equation}
where
\begin{align}
\partial_t f_\kappa = \frac{\partial f_\kappa}{\partial r}\partial_t r + \frac{\partial f_\kappa}{\partial \psi} \partial_t\psi + \frac{\partial f_\kappa}{\partial \psi'} \partial_t \psi' + \frac{\partial f_\kappa}{\partial h} \partial_t h
\end{align}
describes contributions from bending energy density
\begin{equation}
f_\kappa
=
\frac{1}{2}\kappa
\left(
\frac{\psi'}{h}
+
\frac{\sin\psi}{r}
\right)^2,
\end{equation}
and the terms 
\begin{align}
S_1 & = \eta_b \left(\bar{S}_{uu}+\bar{S}_{\phi\phi}\right)^2\\
S_2 & = \eta_s \left[\left(\bar{S}_{uu}-\bar{S}_{\phi\phi}\right)^2+\left(2\bar{S}_{u\phi}\right)^2\right]\\
S_3 & = \frac{1}{\Lambda}\left[r_p+\xi\left(\bar{S}_{uu}+\bar{S}_{\phi\phi}\right)\right]^2    
\end{align}
contain contributions from the strain rate tensor components
 \begin{align}
\bar{S}_{uu} & = \frac{1}{h}\bar{v}_u'+\frac{\psi'}{h}v_n \label{eq:Suu}\\
\bar{S}_{\phi\phi} & = \frac{\cos\psi}{r}\bar{v}_u+\frac{\sin\psi}{r}v_n \label{eq:Spp}\\
\bar{S}_{u\phi} & = \frac{1}{2}\left(\frac{\bar{v}_{\phi}'}{h}-\frac{\cos\psi}{r}\bar{v}_{\phi}\right)\label{eq:Sup}.
 \end{align}
Specific to the present work is the external force $\mathbf{f}^{\text{ext}}_c$ in Eq.~(\ref{eq:FinalRay}), which is given by
\begin{equation}
\mathbf{f}_e^{\mathrm{ext}} = \left(p + f_c^{n}\right)\mathbf{n},
\end{equation}
where the pressure difference $p$ acts as a Lagrange multiplier that enforces
the conservation of the enclosed volume~$V_{\mathrm{cell}}$ the second term is given by
\begin{equation}
\label{eq:fc}
f_c^{n}
=
-\left(
\frac{\partial P^{\mathrm{ext}}}{\partial r}\,\sin\psi
+
\frac{\partial P^{\mathrm{ext}}}{\partial z}\,\cos\psi
\right),
\end{equation}
for the confinement potential $P^{\mathrm{ext}}$ written in Eq.~\eqref{eq:confine_potential} (main text).\\

\begin{figure}[H]
    \centering
	\includegraphics[width = 0.5\columnwidth]{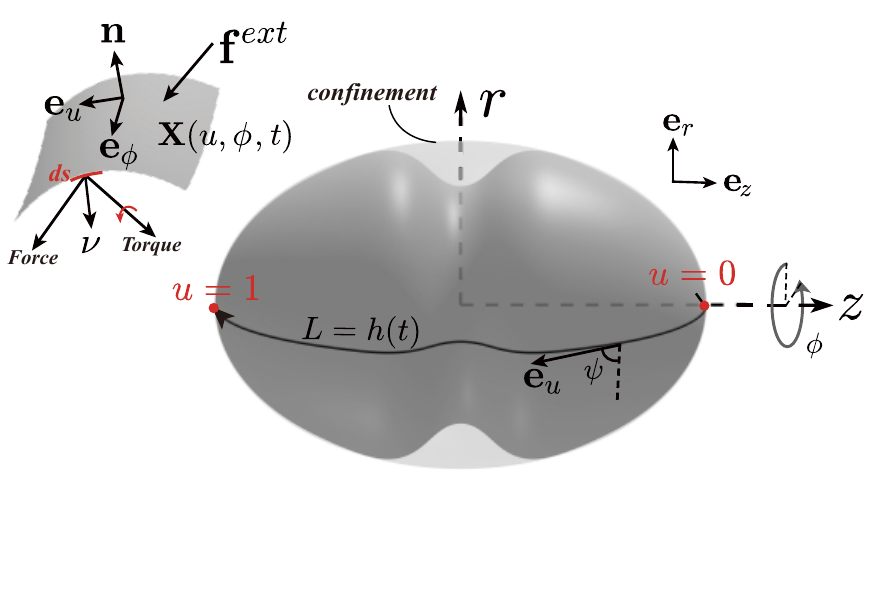}
	\caption{\textbf{Parameterization of the closed axisymmetric surface $\mathbf{X}(u,\phi,t)$.} The transparent gray prolate surface represents the geometric confinement. The coordinates $u=0$ and $u=1$ correspond to the two poles of the surface, respectively. Under the SLE parameterization, the metric factor $h(u,t)=h(t)$ is independent of $u$ and equal to the total arclength $L$.} \label{fig:scheme}
\end{figure}

The second line in Eq.~(\ref{eq:FinalRay}) imposes the geometric relationships given by Eqs.~(\ref{eq:dur}) and (\ref{eq:duz}) with Lagrange multipliers $\alpha$ and $\beta$. The Lagrange multiplier $\zeta$ enforces a scaling Lagrangian–Eulerian (SLE) parameterization defined by the condition
\begin{equation}
\label{eq:SLE}
h' = 0,
\end{equation}
a gauge choice for the paramterization that can always be made~\cite{gao2025}. With this choice, the scaling factor becomes spatially uniform, $h(u,t)=h(t)$ and only depends on time. Consequently, the map between mesh coordinate $u$ and physical arc length $s$ along the meridional contour is simply
\begin{equation}
s = h(t)\,u.
\end{equation}
Accordingly, $h(t)$ coincides with the total arclength $L$ of the meridional outline (Fig.~\ref{fig:scheme}).\\

In ref.~\cite{gao2025}, we have shown that 
\begin{equation}
\label{eq:Funcvar}
\frac{\delta\mathcal{R}}{\delta \Phi} = 0 .
\end{equation}
with $\mathcal{R}$ given in Eq.~(\ref{eq:FinalRay}) and
\[
\Phi \in
\left\{
v^u,\,
v^\phi,\,
\partial_t r,\,
\partial_t z,\,
\partial_t h,\,
\partial_t \psi,\,
\partial_t \alpha,\\,
\partial_t \beta,\,
\partial_t \zeta
\right\},
\]
yields a system of equations that is equivalent to the force and moment balance equations of the active fluid surface. Boundary terms of this variation provide boundary conditions (see Tab.~\ref{Table:Boundry}) for solving the resulting boundary value problem.

\subsection{Boundary value problems}
From the procedure described above, we find dynamic equations for $\partial_t r$, $\partial_t z$, and $\partial_t \psi$. Depending on the need for dynamic simulations or the computation of numerically exact stationary surfaces, we consider two strategies:
\begin{itemize}
    \item To perform dynamic simulations, we connect two time steps that are separated by $\Delta t$ by setting the time derivatives to
\begin{equation}
\begin{split}
    \partial_t r & = \frac{r(u,t)-r(u,t-\Delta t)}{\Delta t}\\
    \partial_t z & = \frac{z(u,t)-z(u,t-\Delta t)}{\Delta t}\\
    \partial_t \psi & = \frac{\psi(u,t)-\psi(u,t-\Delta t)}{\Delta t},\\
\end{split}
\end{equation}
and using $r(u,t-\Delta t), z(u,t-\Delta t), \psi(u,t-\Delta t)$ from the previous time-step. This represents a robust implicit integration the dynamic equations. Solutions $[r(u,t),z(u,t),\psi(u,t),v_u,v_{\phi}]$ derived from this boundary value problem directly correspond to the surface geometries and flows at the next time point. To enforce volume conservation, we introduce the auxiliary volume function $V(u)$, which
satisfies
\begin{equation}
\label{eq:volume}
    V' = \pi r^2 h \sin\psi ,
\end{equation}
together with the boundary conditions
\begin{equation}
\label{eq:bc_volume}
    V(0) = 0, \qquad
    V(1) = V_{\mathrm{cell}} = \frac{4\pi}{3} R_0^3 .
\end{equation}
We solve the boundary value problem arising at every time step using standard solvers as implemented in MATLAB~\cite{bvpsolv01}. The parameter values used in the simulations are listed in Tab.~\ref{tab:para_surface}.
\item To compute steady-state solutions directly, we impose
\begin{equation}
\begin{split}
    \partial_t r &= 0,\\
    \partial_t z &= v_0,\\
    \partial_t \psi &= 0,
\end{split}
\end{equation}
in the system of equations arising from the variation Eq.~(\ref{eq:Funcvar}), where $v_0$ is an additional parameter representing a possible translational
velocity of the surface along the symmetry axis. In the absence of confinement,
$v_0$ is an unknown that must be determined as part of the boundary value
problem. An additional boundary condition $z(0) + z(1) = 0$ is used to fix the reference frame that co-moves with the surface~\cite{Mietke2019Self}. In the presence of confinement, steady state requires $v_0 = 0$. We therefore introduce an additional boundary condition
\begin{equation}
    z(0) + z(1) = \Delta z ,
\end{equation}
and tune the parameter $\Delta z$ such that the solution satisfies $v_0 = 0$.

Steady-state solutions are obtained iteratively by gradually increasing the activity parameter $\xi$ (or equivalently the Péclet number $\Pe$), using the solution at the previous parameter value as the initial guess. In the presence of degeneracies, where multiple steady-state solutions exist for a given $\Pe$, continuation through turning points is achieved by instead incrementing the pole-to-pole distance
\begin{equation}
    H = z(0) - z(1),
\end{equation}
while treating the contractility $\xi$ as an additional unknown parameter.
\end{itemize}

\subsection{Initial conditions}
The dynamic simulations are initialized from a steady state in the absence of contractility ($\Pe = 0$). To obtain this reference state, we begin with a very weak confinement in a prolate geometry, for which a spherical surface with a uniform distribution of stress regulators constitutes a steady-state solution. We then iteratively decrease the confinement dimensions $a$ and $b$, while keeping their ratio fixed at $a/b = 1.3$. This yields deformed, but homogeneous steady-state solutions at progressively stronger confinement that serve as initial conditions for simulations in which activity is turned on.

\subsection{Numerical test of stability of stationary solutions}\label{sec:stabtest}
\begin{figure}[H]
    \centering
	\includegraphics[width = 1\columnwidth]{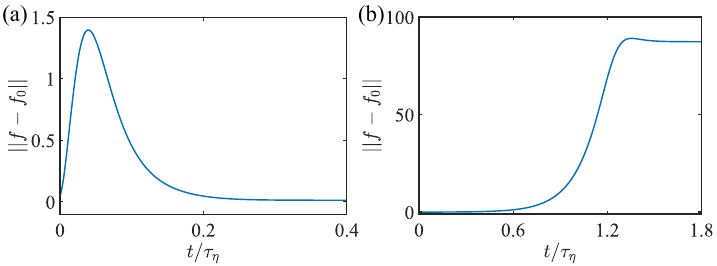}
\caption{\textbf{Stability testing of nonlinear stationary solutions.} Stationary solutions are dynamically simulated with small concentration perturbations as initial conditions. We then measure the time evolution of $||f-f_0||:=||r(u,t)-r_0(u)||_2+||z(u,t)-r_0(u)||_2+||c(u,t)-c_0(u)||_2$, where $||\cdot||_2$ denotes the $L_2$ norm on $u\in[0,1]$. This test typically leads to an unambiguous converging (a) or diverging  dynamics (b), from which we conclude the corresponding point of the branch to be stable or unstable, respectively. Curves show exemplary responses from the stable and unstable parameter regions in Fig.~3b-d.}\label{fig:stabtest}
\end{figure}

Once stationary branches are identified, we test the stability of solutions along these branches by applying a small perturbation $\delta c$ to the steady concentration field $c_0$, i.e. we use $c(u,t=0) = c_0+\delta c$ as an initial condition in a dynamic simulation. The distance $||f-f_0||:=||r(u,t)-r_0(u)||_2+||z(u,t)-r_0(u)||_2+||c(u,t)-c_0(u)||_2$ is used as a measure of the deviation of the dynamic solution $f$ from the steady stationary solution $f_0$. Here, $||\cdot||_2$ denotes the $L_2$ norm on $u\in[0,1]$ (see~Fig.~\ref{fig:stabtest}). 

\section{Linear stability analysis of nonlinear steady states}
Here, we perform a linear stability analysis of weakly ingressed steady states (see red curves in Fig.~\ref{fig:PhaseCurve}a,b,c of the main text). This analysis provides an independent and complementary criterion for assessing the stability of nonequilibrium steady states, alongside direct numerical integration of the full dynamical equations with perturbed steady state taken as the initial condition (see Sec.~\ref{sec:stabtest}). The final result is the semi-analytical expression given in Eq.~(\ref{eq:growthrate}) (main text) for the critical conditions at which spontaneous symmetry breaking occurs.

\subsection{Axial parameterization}
For this analysis, we introduce an axial parameterization
\begin{equation}\label{eq:Axipar}
\mathbf{X}(z,\phi,t)=\bigl[r(z,t)\cos\phi,\; r(z,t)\sin\phi, z\bigr],
\end{equation}
which encodes the surface geometry by a single function $r(z,t)$. While this parameterization is not suitable to describe arbitrary axisymmetric surfaces for which $r(z,t)$ can be multi-valued, it is sufficient for a linear analysis near weakly ingressed shapes that are described by a well-defined single-valued radial function.

For the parameterization Eq.~(\ref{eq:Axipar}), the tangent basis and normal vectors are 
\begin{equation}
\mathbf{E}_z =\sqrt{1+(r')^2}\bar{\mathbf{E}}_z=\begin{pmatrix}
r'\cos\phi\\ r'\sin\phi\\ 1    
\end{pmatrix}
\qquad
\mathbf{E}_\phi = r\bar{\mathbf{E}}_\phi=\begin{pmatrix}
-r\sin\phi\\ r\cos\phi\\ 0    
\end{pmatrix}
\qquad
\mathbf{N}=\frac{1}{\sqrt{1+(r')^2}}
\begin{pmatrix}
-\cos\phi\\ -\sin\phi\\ r'  
\end{pmatrix}.
\end{equation}
where $\bar{\mathbf{E}}_i$ depict normalized tangent vectors. The non-vanishing metric tensor and curvature tensor components are
\begin{equation}
    g_{zz}=\sqrt{1+(r')^2} \qquad g_{\phi\phi}=r^2.
\end{equation}
and
\begin{equation}
    C_z^z=-\frac{r''}{(1+r'^2)^{3/2}} \qquad C^{\phi}_{\phi}=\frac{1}{r\sqrt{1+r'^2}},
\end{equation}
respectively. Here and in the following derivation, primes $'$ denote derivatives with respect to
$z$, and dots $\dot{\ }$ denotes time derivatives.

\subsection{Kinematic equations}
The center of mass flow $\mathbf{v}$ is decomposed as
\begin{equation}
    \mathbf{v}(z,\phi,t)
    =
    v(z,t)\,\bar{\mathbf{E}}_z
    + v_n(z,t)\,\mathbf{n},
\end{equation}
where $v(z,t)$ is the tangential velocity along the meridional direction, and
$v_n(z,t)$ is the normal velocity. The surface dynamics reads
\begin{equation}\label{eq:Xdyn}
    \frac{\mathrm{d}}{\mathrm{d}t}[\mathbf{X}(z,\phi,t)]  = \mathbf{v}(z,\phi,t),
\end{equation}
where the material derivative of a function $f(z,t)$ is defined as
\begin{equation}
\label{eq:materialderivative_z}
    \frac{\mathrm{d}f}{\mathrm{d}t}
    =
    \dot{f}
    + q^z f'.
\end{equation}
Here, $q^z$ denotes the coordinate flow associated with the
parameterization Eq.~(\ref{eq:Axipar}). From Eq.~(\ref{eq:Xdyn}), we find
\begin{equation}
    v_n(z,t)
    =
    \frac{\dot{r}}{\sqrt{1+(r')^2}},
\end{equation}
while the coordinate flow is given by
\begin{equation}
\label{eq:meshqz}
    q^z
    =
    -\frac{\dot{r}\,r'}{1+(r')^2}
    + \frac{v}{\sqrt{1+(r')^2}}.
\end{equation}
\subsection{Mapping between the $\{z,\phi\}$ and $\{u,\phi\}$ parameterizations}
Using the geometric relations
\begin{equation}
    ds = h\,du = \frac{dr}{\cos\psi} = -\frac{dz}{\sin\psi},
\end{equation}
as well as
\begin{equation}
    \cos[\psi(u,t)]
    =
    -\,\frac{\partial_z r}{\sqrt{1+(\partial_z r)^2}}, \quad\quad \sin[\psi(u,t)]
    =
    \frac{1}{\sqrt{1+(\partial_z r)^2}},
\end{equation}
together with the identity $\bar{\mathbf{E}}_z = -\bar{\mathbf{e}}_u$, we
establish the mapping between fields defined in the $\{z,\phi\}$ and
$\{u,\phi\}$ parameterizations. Specifically, we obtain
\begin{align}
r(z,t) &= r(u,t), 
&\partial_z r(z,t)
&= \frac{\partial_u r(u,t)}{\partial_u z(u,t)}
= -\cot\psi(u,t), \\[0.5em]
v(z,t) &= -\bar{v}_u(u,t),
&\partial_z v(z,t)
&= -\frac{\partial_u \bar{v}_u(u,t)}{\partial_u z(u,t)}
= \frac{1}{\sin\psi(u,t)}\,\frac{\partial_u \bar{v}_u(u,t)}{h(u,t)}, \\[0.5em]
c(z,t) &= c(u,t),
&\partial_z c(z,t)
&= \frac{\partial_u c(u,t)}{\partial_u z(u,t)}
= -\frac{1}{\sin\psi(u,t)}\,\frac{\partial_u c(u,t)}{h(u,t)} .
\end{align}
When converting a function $f(z,t)$ to its representation $f(u,t)$, the
corresponding coordinate $u$ is obtained by inverting the relation
$z = z(u,t)$. Finally, with the above identities, it's easy to verify that the coordinate flow Eq.~(\ref{eq:meshqz})
is equivalent to the coordinate flow equation~\cite{gao2025}
\begin{equation}
q^u=v^u+h^{-1}\left(\sin\psi\partial_tz-\cos\psi\partial_tr\right)    
\end{equation}
for the $\{u,\phi\}$ parameterization.
\subsection{Force balance equations}
The force balance equations in the $\{z,\phi\}$ parameterization can be derived
by constructing a Rayleigh dissipation functional analogous to that used in the
$\{u,\phi\}$ coordinate system, and subsequently taking variations with respect
to $\partial_t r(z,t)$ and $v(z,t)$. For brevity, we omit the intermediate steps
of this derivation and present only the resulting equations below. The normal force balance equation reads in axial coordinates
\begin{equation}
\begin{aligned}
\label{eq:fbalance_norm}
D_1 \dot{r} &= \frac{1}{2}{\kappa} \Bigg\{\left[1+(r')^2\right]^3\left[1+2(r')^2\right] - r r'' \left[1 - 3(r')^4 - 2(r')^6\right] \\
&\quad - r^2\left[1+(r')^2\right] \left(3(r'')^2\left[1-4(r')^2\right] + 4 r' r''' \left[1+(r')^2\right]\right) \\
&\quad + r^3 \Big( 5(r'')^3(1-6(r')^2) + 20 r' r'' r''' (1+(r')^2) - 2 r'''' (1+(r')^2)^2 \Big) \Bigg\} \\
&+ {\eta_b} \Bigg\{ -r \left(v r\right)'\left[1+(r')^2\right]^{5/2}\left[1+(r')^2-r r''\right]\Bigg\} \\
&+ {\eta_s} \Bigg\{ - r\left(v r'-rv'\right)\left[1+\left(r'\right)^2\right]^{5/2}\left[1+\left(r'\right)^2+r r''\right] \Bigg\}\\
&+ {\xi}H_+(c)\left[1+(r')^2\right]^3 \Big\{r^3 r'' - r^2\left[1+(r')^2\right]\Big\} + r^3\left[1+(r')^2\right]^{9/2}f_n^{\mathrm{ext}},
\end{aligned}
\end{equation}
where
\begin{equation}
D_1 = r\left[1+(r')^2\right] \Bigg\{(\eta_s + \eta_b) \left[(1+(r')^2)^2 + r^2 (r'')^2 \right] + 2(\eta_s - \eta_b) r\left[1+(r')^2\right] r'' \Bigg\}.
\end{equation}

The meridional force balance equation reads in axial coordinates:
\begin{equation}
\label{eq:fbalance_tan}
\begin{split}
D_2\dot{r'} &={\eta_s} \Bigg\{ - \dot{r} r'\left[1+(r')^2\right]^{5/2} - v (r')^2 \left[1+(r')^2\right]^3 + v' r r' \left[1+(r')^2\right]^3 \\
&\quad - v r r'' \left[1+(r')^2\right]^2 + v'' r^2 \left[1+(r')^2\right]^3 - v' r^2 r' r'' \left[1+(r')^2\right]^2 \\
&\quad - \dot{r} r^2 r''' \left[1+(r')^2\right]^{3/2} + 4 \dot{r} r^2 r' (r'')^2 \sqrt{1+(r')^2} \Bigg\}\\
&+ {\eta_b} \Bigg\{- \dot{r} r'\left[1+(r')^2\right]^{5/2} - v (r')^2 \left[1+(r')^2\right]^3 + v' r r' \left[1+(r')^2\right]^3 \\
&+ v r r'' \left[1+(r')^2\right]^2 + v'' r^2 \left[1+(r')^2\right]^3 - v' r^2 r' r''\left[1+(r')^2\right]^2 \\
&- \dot{r} r^2 r'''\left[1+(r')^2\right]^{3/2} + 4 \dot{r} r^2 r' (r'')^2 \sqrt{1+(r')^2} - 2 \dot{r} r r' r''\left[1+(r')^2\right]^{3/2} \Bigg\}\\
&+{\xi} r^2 c' g'(c) \left[1+(r')^2\right]^{7/2}-{\Gamma}v r^2 \left[1+(r')^2\right]^4,
\end{split}
\end{equation}
where 
\begin{equation}
D_2 = r \left[1+(r')^2\right]^{3/2} \Bigg\{\eta_s \Big[1+(r')^2 + r r'' \Big] + \eta_b \Big[ r r'' - [1+(r')^2] \Big] \Bigg\}.
\end{equation}

\subsection{Dynamic equation of the concentration field}
Using the definition of the material derivative in
Eqs.~\eqref{eq:materialderivative_z} and \eqref{eq:meshqz}, the dynamics Eq.~\eqref{eq:dcdt} governing the evolution of the concentration field $c$ in the
$\{z,\phi\}$ parameterization reads
\begin{equation}
\label{eq:dcdt_z}
\dot{c} = {\dot{r}}\frac{\left[1+(r')^2\right]\left(r r' c'-c\right)+r r'' c}{r\left[1+(r')^2\right]^2}+ {D}\frac{-r r' r'' c' + (r' c'+r c'')\left[1+(r')^2\right]}{r\left[1+(r')^2\right]^2} - k\left(c-\hat{c}_{0} \right) - \frac{(vcr)'}{r\sqrt{1+(r')^2}}.
\end{equation}

\subsection{Linearized equations}
We assume that the force balance equations
Eqs.~\eqref{eq:fbalance_norm} and \eqref{eq:fbalance_tan},
together with the concentration dynamics
Eq.~\eqref{eq:dcdt_z}, admit a steady-state solution
$\{r_{ss}(z),\, v_{ss}(z),\, c_{ss}(z)\}$.
These steady states correspond to a symmetric, partially ingressed geometry, characterized by cortical flows directed from the poles toward the equator and by an accumulation of stress regulators at the equatorial region, located at $z = 0$. Consequently, the steady-state profiles $r_{ss}(z)$ and $c_{ss}(z)$ are even functions of $z$, whereas the flow profile $v_{ss}(z)$ is an odd function of
$z$. Therefore, we must have at the symmetry plane $z = 0$:
\begin{equation}
\begin{split}
    r_{s0}' &\equiv r_{ss}'(0) = 0,
    \qquad
    r_{s0}''' \equiv r_{ss}'''(0) = 0, \\
    c_{s0}' &\equiv c_{ss}'(0) = 0, \\
    v_{s0} &\equiv v_{ss}(0) = 0,
    \qquad
    v_{s0}'' \equiv v_{ss}''(0) = 0 .
\end{split}
\end{equation}

We then introduce a small perturbation around this steady state,
\begin{equation}
\label{eq:symmetry}
\begin{split}
    r(z,t) &= r_{ss}(z) + \epsilon\,\delta r(z,t), \\
    v(z,t) &= v_{ss}(z) + \epsilon\,\delta v(z,t), \\
    c(z,t) &= c_{ss}(z) + \epsilon\,\delta c(z,t),
\end{split}
\end{equation}
and expand the governing equations to first order in the small parameter $\epsilon$. The zeroth-order terms reproduce the steady-state equations, while the terms to $\mathcal{O}(\epsilon)$ yield a set of linearized evolution equations for the perturbations $\delta r(z,t)$, $\delta v(z,t)$, and $\delta c(z,t)$. The latter can be cast into the compact form
\begin{equation}
\label{eq:linearized_system}
    L
    \begin{pmatrix}
        \dot{\delta r} \\
        \dot{\delta c} \\
        0
    \end{pmatrix}
    =
    M
    \begin{pmatrix}
        \delta r \\
        \delta c \\
        \delta v
    \end{pmatrix},
\end{equation}
where $L$ and $M$ are linear differential operators whose coefficients depend on the steady-state profiles $r_{ss}(z)$, $v_{ss}(z)$, and $c_{ss}(z)$, as well as their derivatives.

The linear stability of the steady state is determined by the generalized eigenvalue problem
\begin{equation}
\label{eq:eigen_problem}
    Lf_n = \lambda_nMf_n,
\end{equation}
where $f_n$ are suitable eigenmodes that are unstable if their eigenvalue
$\lambda_n$ has a positive real part. Equation~(\ref{eq:eigen_problem}) is in general analytically intractable and must therefore be solved
numerically. 

To make progress, we assume that the onset of symmetry breaking is dominated by perturbations localized in the vicinity of the equatorial plane $z = 0$. Accordingly, we approximate the linear differential operators $L$ and $M$ by evaluating their coefficients at $z = 0$. Exploiting the symmetry properties of the steady state summarized in Eq.~(\ref{eq:symmetry}), the normal force balance equation Eq.~\eqref{eq:fbalance_norm} simplifies to
\begin{equation}
\label{eq:fbalance_norm_pert}
\begin{aligned}
D_{1,0}\dot{\delta r} &= \frac{1}{2}{\kappa} \Bigg\{r_{s0}^2 \left(3 r_{s0} r_{s0}'' \left(5 r_{s0} r_{s0}''-2\right)-1\right)
   \delta r''-2 r_{s0}^4 \delta r''''+\left(r_{s0}^3 \left(5 (r_{s0}'')^3-2
   r_{s0}''''\right)+r_{s0} r_{s0}''-2\right) \delta r\Bigg\} \\
&+ {\eta_b}r_{s0}^3 \Big[\delta v' \left(r_{s0} r_{s0}''-1\right)+v_{s0}' \left(r_{s0} \delta r''+r_{s0}'' \delta r \right)\Big] + {\eta_s}r_{s0}^3 \Big[\delta v' \left(r_{s0} r_{s0}''+1\right)+v_{s0}' \left(r_{s0} \delta r''+r_{s0}''\delta r\right)\Big] \\
&+ {\xi} r_{s0}^3 \Big[H_+\left(c _{s0}\right) \left(r_{s0}'' \delta r+r_{s0}
   \delta r''\right)+\left(r_{s0} r_{s0}''-1\right)
   H_+'\left(c_{s0}\right)\delta c\Big] 
   +  {f_n^{\text{ext}}}r_{s0}^3 \delta r,
\end{aligned}
\end{equation}
where
\begin{equation}
D_{1,0}= r_{s0}^2 \big[\eta_s \left(r_{s0} r_{s0}''+1\right){}^2+\eta_b \left(r_{s0}r_{s0}''-1\right){}^2\big].
\end{equation}
Similarly, the meridional force balance equation
Eq.~\eqref{eq:fbalance_tan} reduces to
\begin{equation}
\label{eq:fbalance_tan_pert}
\begin{aligned}
D_{2,0} \dot{\delta r'} &={\eta_s} \Bigg[ - r_{s0}'' \delta v + v_{s0}' \delta r' \Big( 1 - r_{s0} r_{s0}'' \Big) + r_{s0} \delta v'' \Bigg]+ {\eta_b} \Bigg[ r_{s0}'' \delta v + v_{s0}' \delta r' \Big( 1 - r_{s0} r_{s0}'' \Big) + r_{s0} \delta v'' \Bigg]\\
&+ {\xi}r_{s0} H_+'(c_{s0}) \delta c'-{\Gamma} r_{s0} \delta v  
\end{aligned}
\end{equation}
where
\begin{equation}
D_{2,0} = \eta_s-\eta_b+ (\eta_s+\eta_b) r_{s0} r_{s0}'' .
\end{equation}
Finally, the concentration dynamics equation Eq.~\eqref{eq:dcdt_z} yields to linear order
\begin{equation}
\label{eq:ConDynamics_pert}
\delta \dot{c}+ \left( v_{s0}' \delta c + c_{s0} \delta v' \right) = {\dot{\delta r}} \left[ c_{s0} \left( r_{s0}'' - \frac{1}{r_{s0}} \right) \right] + {D}  \delta c'' - {k}  \delta c  \\
\end{equation}

We next seek normal-mode solutions of the form
\begin{equation}\label{eq:NMansatz}
\begin{split}
\delta r(z,t) &= \delta r(t)\, e^{iq z},\\
\delta v(z,t) &= \delta v(t)\, e^{iq z},\\
\delta c(z,t) &= \delta c(t)\, e^{iq z},
\end{split}
\end{equation}
where $q$ is the perturbation wavenumber.
Substituting these expressions into
Eqs.~\eqref{eq:fbalance_norm_pert}, \eqref{eq:fbalance_tan_pert}, \eqref{eq:ConDynamics_pert} yields a closed system of linear ordinary differential equations for the time-dependent coefficients $\delta r(t),\,\delta v(t)$ and $\delta c(t)$ in Eqs.~(\ref{eq:NMansatz}). This system can be written as

\begin{equation}
    \begin{pmatrix}
    \dot{\delta r} \\
    \dot{\delta c}
    \end{pmatrix} = 
    \begin{pmatrix}
    J_{rr} & J_{rc}\\
    J_{c r} & J_{cc}
    \end{pmatrix} \begin{pmatrix}
    \delta r \\
    \delta c
    \end{pmatrix}
\end{equation}
where the components of the $2\times 2$ Jacobian matrix are given by
\begin{align}
J_{rr} &= {\xi}H_+(c_{s0})\frac{ (r_{s0}'' - q^2 r_{s0})}{2 \eta_b r_{s0}}+ r_{s0}^2 v_{s0}' \left[(r_{s0}'')^2 - q^2\right]  +\frac{f^{\text{ext}}_n}{2 \eta_b r_{s0}}\notag\\
&+ {\kappa}\frac{-2 + r_{s0} \left\{ r_{s0}'' + r_{s0} \left[ -2 q^4 r_{s0}^2 + q^2 [1 + 3 r_{s0} r_{s0}'' (2 - 5 r_{s0} r_{s0}'')] + r_{s0} [5 (r_{s0}'')^3 - 2 r_{s0}^{''''}] \right]\right\}}{4 \eta_b r_{s0}^2}\label{eq:Jrr},\\
J_{rc} &= -{\xi} \frac{r_{s0}H_+'(c_{s0})}{2\eta_b}\label{eq:Jrc},\\
J_{c r} &= {\xi} H_+(c_{s0})\frac{c_{s0}  (q^2 r_{s0} - r_{s0}'')}{2 \eta_b} + \frac{c_{s0} v_{s0}' \Big\{1 + r_{s0} \big[ -r_{s0}'' + r_{s0} (q^2 - (r_{s0}'')^2) \big] \Big\}}{r_{s0}}  -  \frac{c_{s0}f_n^{\text{ext}}}{2 \eta_b} \notag\\
&+ {\kappa}\frac{c_{s0} \Big[ 2 - q^2 r_{s0}^2 + 2 q^4 r_{s0}^4 - r_{s0} r_{s0}'' - 5 r_{s0}^3 (r_{s0}'')^3 + 2 r_{s0}^3 r_{s0}^{(4)} + 3 q^2 r_{s0}^3 r_{s0}'' (5 r_{s0} r_{s0}'' - 2) \Big]}{4 \eta_b r_{s0}^3},\\
J_{cc} &= 
-{k} 
- {D}q^2 
+ {\xi} \frac{c_{s0} H_+'(c_{s0})}{\eta_b} 
- v_{s0}'.\label{eq:Jcc}
\end{align}
In Eqs.~\eqref{eq:Jrr}–\eqref{eq:Jcc}, we have set $\eta_s=\eta_b$ to simplify expressions.\\

\begin{figure}[H]
	\includegraphics[width = 1\columnwidth]{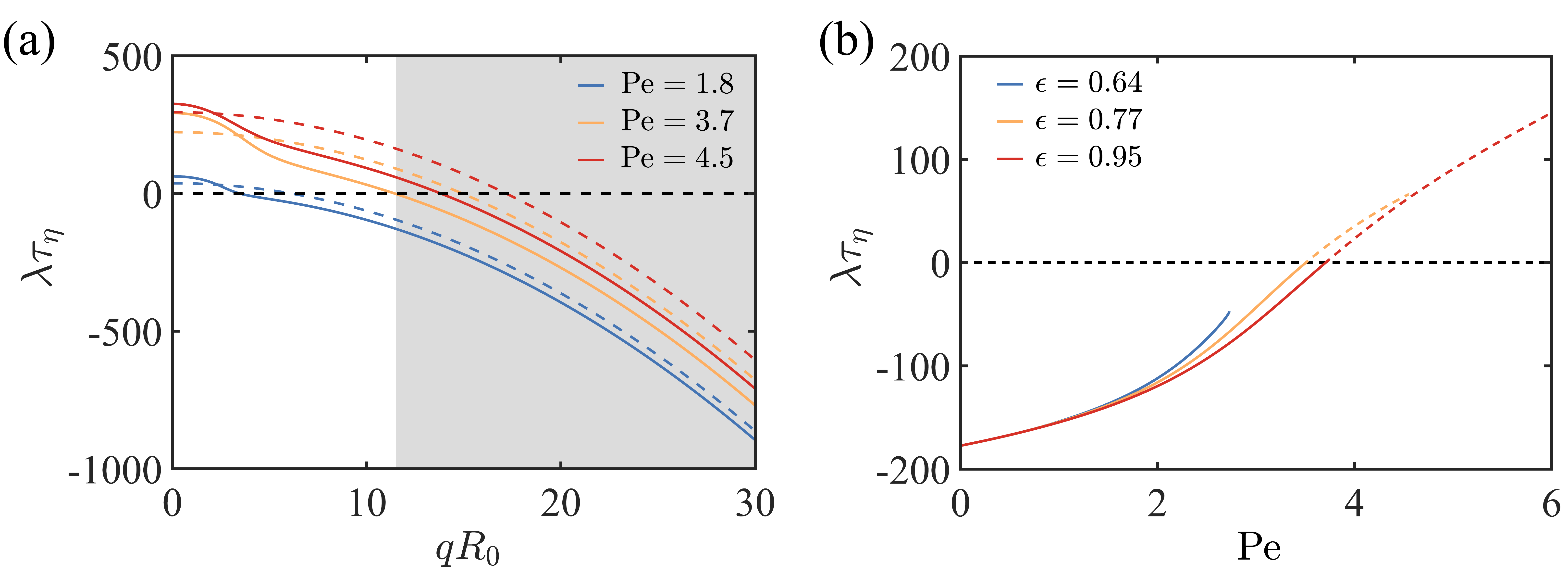}
	\caption{\textbf{Exemplary linear stability analysis of nonequilibrium steady states under strong confinement ($\boldsymbol{\epsilon = 0.946}$)}. (a)~Growth rate $\lambda$ as a function of the dimensionless wave number $qR_0$. Only modes within the shaded region ($q > q_{\mathrm{min}} = 11.5/R_0$) are considered. Solid curves show the largest real part of the eigenvalues of the Jacobian matrix $J$, while dashed curves show its $J_{cc}$ component. (b)~Growth rates as a function of $\mathrm{Pe}$ for different confinement degrees as predicted from the Jacobian with components Eqs.~(\ref{eq:Jrr})--(\ref{eq:Jcc}). To obtain these curves, the Jacobian is computed using the steady state values (and derivatives) of the radial coordinate $r_{s0}$. the concentration $c_{s0}$ and flows $v_{s0}$ along the branch of symmetric, weakly ingressed surfaces (Fig.~\ref{fig:PhaseCurve}b--d, red curves) The change of the growth rate from negative to positive values indicates the onset of an instability. Solid and dashed segments of the curves denote stable and unstable regimes, respectively, as determined from direct numerical integration of the full dynamical equations, indicated very good agreement with the semi-analytic stability analysis. \label{fig:growthrate}}
\end{figure}

As we assume that stability is primarily governed by the local behavior of the perturbation equations in the vicinity of the equatorial plane $z = 0$, corresponding to short-wavelength modes, we restrict our analysis to perturbations with wave numbers $q$ larger than a minimal value $q_{\mathrm{min}}$. For three representative values of contractility, Fig.~\ref{fig:growthrate}a shows the growth rate -- defined as the real part of the largest eigenvalue of the Jacobian matrix $J$ with components Eqs.~(\ref{eq:Jrr})--(\ref{eq:Jcc}) -- as a function of the wave number $q$. The values for $c_{s0},\,r_{s0}$, and $v_{s0}$, as well as their derivatives, are taken from numerical steady state solutions along the weakly ingressed solution branch. The growth rate predicted by the Jacobian above decreases monotonically with increasing $q$. Consequently, if the growth rate is negative at $q=q_{\mathrm{min}}$, it remains negative for all $q>q_{\mathrm{min}}$, and the steady state is linearly stable. Conversely, if the growth rate is positive at $q=q_{\mathrm{min}}$, the steady state is unstable. We therefore use the value of the largest eigenvalue of the Jacobian matrix evaluated at $q=q_{\mathrm{min}}$ as a criterion for assessing the stability of the steady state.\\

Given a prescribed value of $q_{\mathrm{min}}$, we can predict the critical P\'eclet number $\mathrm{Pe}^*$ at which the growth rate $\lambda(q_{\mathrm{min}})$ acquires a positive real part. We treat $q_{\mathrm{min}}$ as a single fitting parameter and determine its optimal value by minimizing the discrepancy between the critical P\'eclet numbers $\mathrm{Pe}^*_a$ obtained from the semi-analytical linear stability analysis and $\mathrm{Pe}^*_n$ determined from direct numerical simulations across different degrees of confinement. In practice, this corresponds to minimizing the distance between the points marked by $\circ$ and $+$ in Fig.~\ref{fig:PhaseCurve}a of the main text. The growth rates computed from the semi-analytical eigenvalue analysis are shown in Fig.~\ref{fig:growthrate}b. Notably, with $q_{\mathrm{min}}R_0=11.5$, they become positive at almost the same P\'eclet numbers at which the direct numerical simulations predict the onset of shape instabilities, indicated by the transition from solid to dashed lines.\\

Heuristically, we find that the growth rates predicted by the Jacobian are well approximated by the matrix element~$J_{cc}$ given in Eq.~\eqref{eq:Jcc} (compare solid and dashed lines in Fig.~\ref{fig:growthrate}a), which motivates the discussion of Eq.~(\ref{eq:growthrate}) in the main text. In Eq.~\eqref{eq:Jcc}, both the turnover and diffusion terms contribute negatively to the growth rate and, therefore, act to stabilize the steady state. The steady-state concentration of stress regulators at the equator, $c_{s0}$, increases monotonically with the contractility $\Pe$. However, the term $c_{s0} H_+'(c_{s0})$ exhibits a nonmonotonic dependence on $c_{s0}$, contributing only a small positive correction at large $c_{s0}$ (or $\mathrm{Pe}$). In contrast, the velocity gradient term $-v_{s0}'$ provides a dominant positive contribution, which ultimately drives the instability.   

\newpage

\section{Supplementary figures}

\begin{figure}[H]
    \centering
	\includegraphics[width = 0.5\columnwidth]{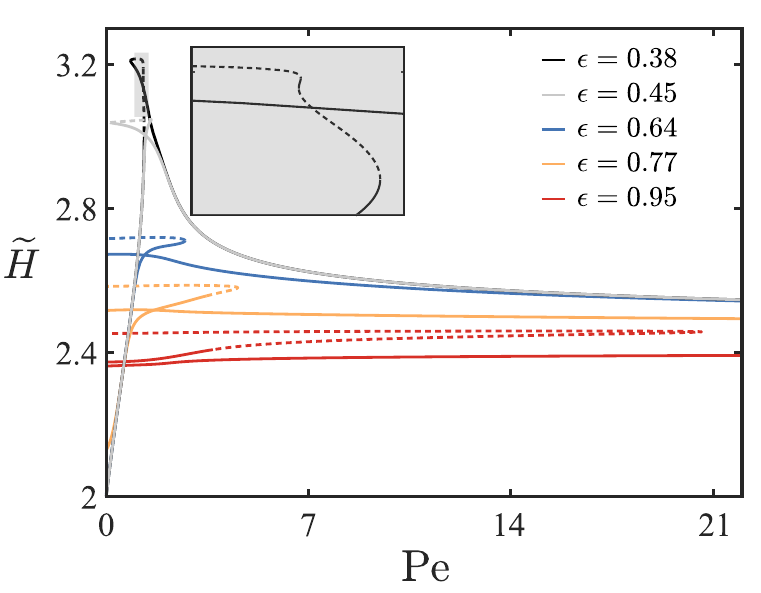}
	\caption{\textbf{Confinement-induced deformation and break-up of the Gibbs loop in the stationary solution space.} At weak confinement ($\epsilon = 0.38,\,0.45$), solution branches form loops that continuously connects solutions corresponding to partially ingressed and fully ingressed geometries (see \cite{gao2025} for corresponding plot in the absence of any confinement). In contrast, at strong confinement ($\epsilon = 0.64,\,0.77,\,0.95$), these branches become disconnected. Insets show a zoomed-in view of the loop structure for $\epsilon = 0.38$. All other parameters are listed in Table~\ref{tab:para_surface}.}\label{fig:top_Tran}
\end{figure}

\begin{figure}[H]
    \centering
	\includegraphics[width = 0.5\columnwidth]{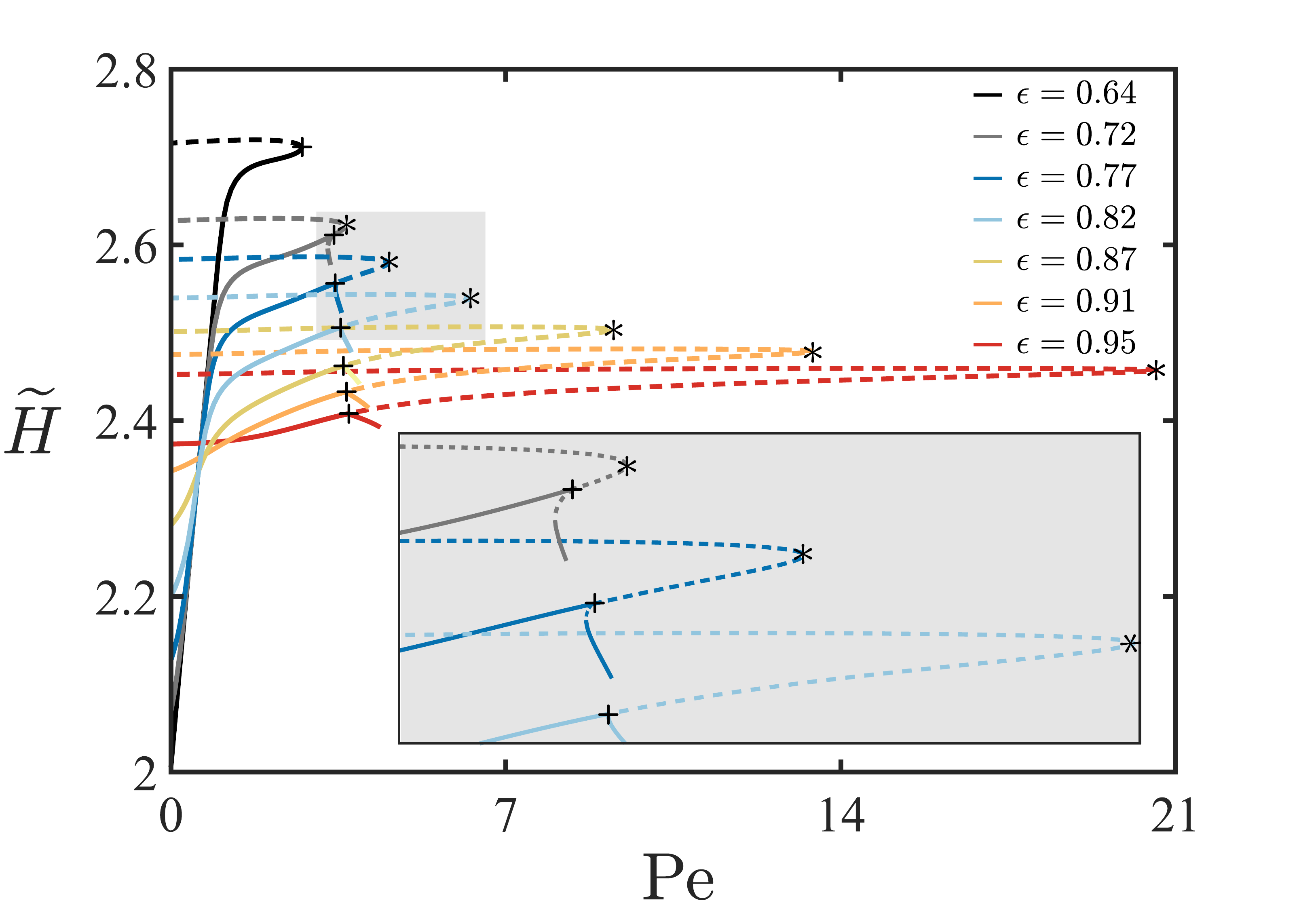}
	\caption{\textbf{Critical points of weakly ingressed solution branches indicate transitions between symmetric and polarized surface shapes.} At weak confinement ($\epsilon= 0.64$), solutions remain stable (solid segment of the black curve) up to the turning point ($*$). As the degree of confinement increases, this branch becomes unstable before reaching the turning point, and a branch of asymmetric, partially ingressed geometries bifurcates from the symmetric branch (bifurcations indicated by $+$, see inset). For P\'eclet numbers beyond turning points, the surface undergoes a strong symmetric ingression. The location of these critical points in the stationary solution space agrees well with the boundaries of regions of different dynamically emerging surface geometries depicted in the phase diagram Fig.~\ref{fig:PhaseCurve}a (main text). All other parameters are listed in Table~\ref{tab:para_surface}.}\label{fig:Pe*_Tran}

\end{figure}

\begin{figure}[H]
	\includegraphics[width = 1\columnwidth]{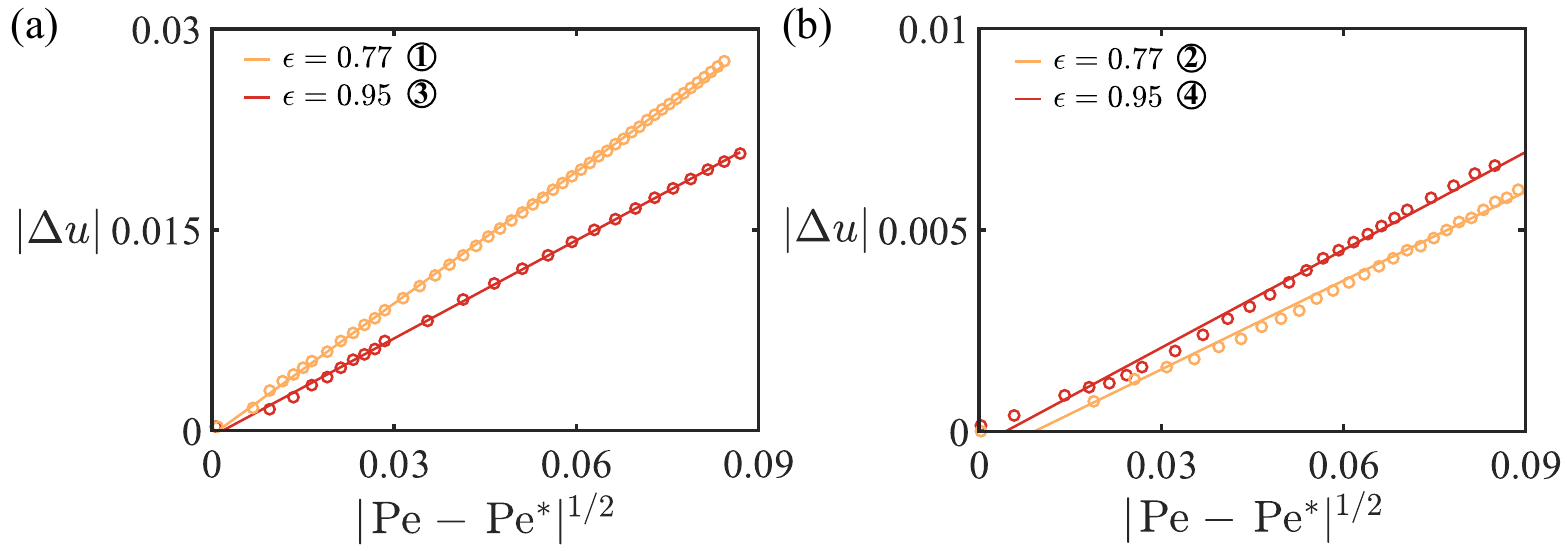}
	\caption{\textbf{Scaling behavior near pitchfork bifurcations.} $|\Delta u| = |u_{\max} - 1/2|$ denotes the deviation of the location of the maximum concentration, $u_{\max}$, from the equatorial plane at $u = 1/2$. $\mathrm{Pe}^*$ denotes critical contractility P\'eclet number of the bifurcation. (a)~Scaling behavior near the bifurcation points labeled \textcircled{1} and \textcircled{3} in Fig.~3e,g (branching of weakly ingressed polarized surface geometries). (b)~Scaling behavior near the bifurcation points labeled \textcircled{2} and \textcircled{4} in Fig.~3f,h (branching of strongly ingressed polarized surface geometries).}\label{fig:Bifur_varAc_Root}
\end{figure}

\begin{figure}[H]
    \centering
	\includegraphics[width = 1\columnwidth]{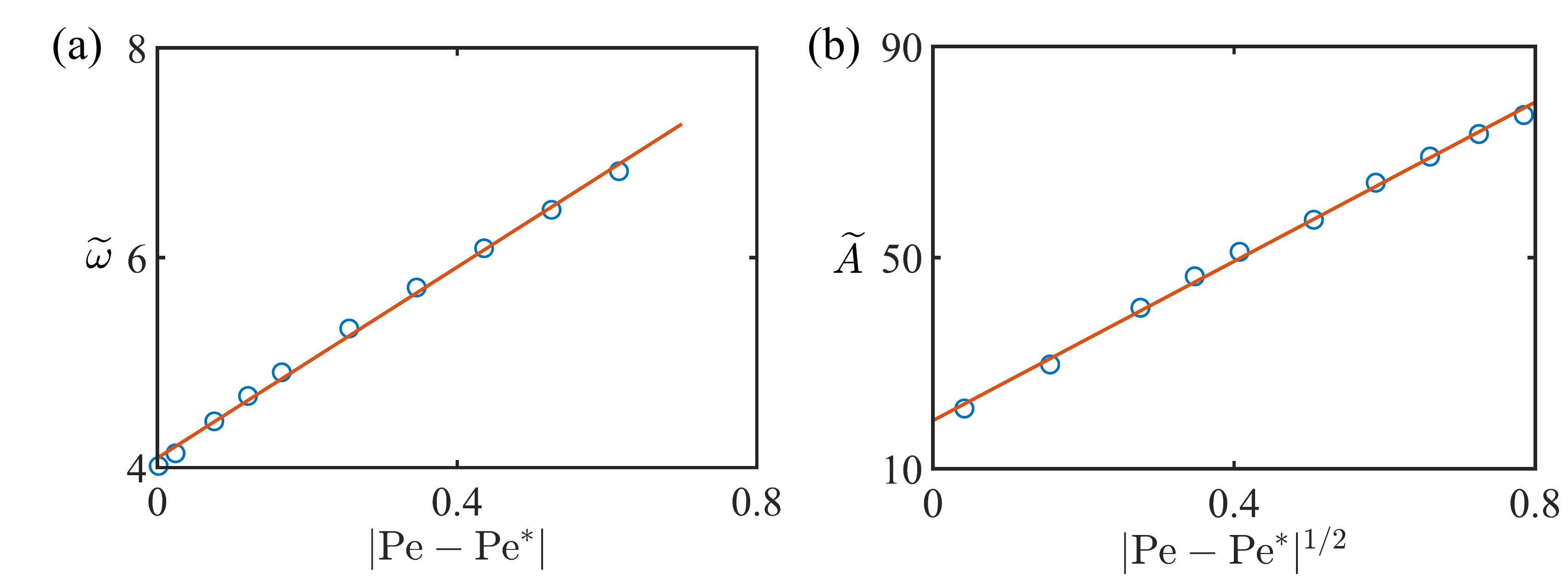}
	\caption{\textbf{Exemplary scaling signatures of a Hopf bifurcation at the onset of polar shape oscillations.} (a)~Oscillation frequency~$\tilde{\omega}=\omega\tau_{\eta}$. (b)~Oscillation amplitude $\tilde{A}=c\left(u=0\right)/c_0$.  Data shown uses confinement magnitude $\epsilon=0.87$ and bifurcation point $\text{Pe}^*=4.24$, all other parameters are listed in Table~\ref{tab:para_surface}. }\label{fig:Oscillation_fit}
\end{figure}

\newpage

\section{Supplementary tables}


\begin{table}[H]
    \centering
    \begin{tabular}{cccc}
    \hline
        \multicolumn{2}{c}{Parameter} & Value & Unit\\
        \hline 
        $\kappa$ & Bending rigidity &  1 & $\kappa$ \\         
        $\eta_b$ & Bulk viscosity & 1  & $\eta_b$   \\
        $\eta_s$ & Shear viscosity & 1 & $\eta_b$  \\      
        $R_0$ & Spherical radius & 1 & $R_0$ \\
        $\tau_\eta$ & Time scale & $\eta_bR_0^2/\kappa$ & $\tau_\eta$\\
        $\Delta t$ & Temporal resolution & 1/900 & $\tau_\eta$ \\
        $\Gamma$ & Friction & 0.09 & $\eta_b/R_0^2$ \\
        $D$ & Diffusion constant & 1 & $R_0^{2}/\tau_\eta$\\
        $k$ & Turnover rate & 45 & $\tau_\eta^{-1}$\\
        $c_0$ & Steady state concentration & 1 & $c_0$ \\
        $c_s$ & Saturation concentration & 10 & $c_0$ \\
       $w$ & Width of the regulated region & 0.14 & 1\\
        $w_p$ & Thickness of the confined shell & 0.14 & 1 \\
        $P_0$ & Strength of confinement & 0.9 & $\kappa/R_0^2$ \\
    \hline
    \end{tabular}
    \caption{Parameters used in closed active fluid surface examples shown in Figs.~\ref{fig:Shell_effect}--\ref{fig:PhaseCurve}. The characteristic time in these simulations is defined as $\tau_{\eta}=\eta_b R_0^2/\kappa$. Characteristic length scale $R_0$ is related to the enclosed conserved volume by $V = 4\pi R_0^3/3$.}
    \label{tab:para_surface}
\end{table}

\begin{table}[H]
    \centering
    \begin{tabular}{cc}
    \hline
        Variation & \ Implied ODE\\         
        \hline
         $\partial_t r$ & $\alpha' = \cdots $ \\
         $\partial_t z$ & $\beta' = \cdots $ \\
         $\partial_t\psi$ & $\psi'' = \cdots$  \\
         $\partial_t h$ &    $\zeta' = \cdots $ \\
         $\partial_t\alpha$ & $r' = \cdots $ \\
         $\partial_t\beta$ & $z' = \cdots $   \\
         $\partial_t \zeta$ & $h' = 0\hspace{0.32cm}$      \\
         $v^u$ & $(v^u)'' = \cdots $ \\
         $v^{\phi}$ & $(v^{\phi})'' = \cdots $\\
    \hline
    \end{tabular}
    \caption{Structure of the ODE system arising from variation described in Sec.~\ref{sec:NumApr}. Further details can be found in~\cite{gao2025}.}
    \label{table:ODE}
\end{table}

\begin{table}[H]
    \centering
    \begin{tabular}{ccc}
    \hline
        \multirow{1}{*}{Variation}  & \multicolumn{2}{c}{\ Boundary conditions}\\ 
        \hline
         $\partial_t r$ & $r(0) = 0,\,r(1) = 0$ \\
         $\partial_t z$ & $\beta(0) = 0,\,\beta(1) = 0$ \\
         $\partial_t\psi$ & $\psi(0) = 0$, $\psi(1) = \pi$ \\
         $\partial_t h$  & $\zeta(0) = 0,\,\zeta(1) = 0$ \\
         $v^u$ & $v^u(0) = 0,\,v^u(1) = 0$ \\
         $v^{\phi}$ & $v^{\phi}(0) = 0,\,v^{\phi}(1) = 0$ \\ 
    \hline
    \end{tabular}
    \caption{Boundary conditions arising from variation described in Sec.~\ref{sec:NumApr}. Further details can be found in~\cite{gao2025}.}
    \label{Table:Boundry}
\end{table}

\bibliography{apssamp} 
\end{document}